\documentclass[11pt,onecolumn,floatfix,altaffilletter,superscriptaddress,
        tightenlines,showpacs,showkeys,preprintnumbers,nofootinbib]{revtex4-1}
\pdfoutput=1
\usepackage[colorlinks=true,citecolor=blue,linkcolor=blue,breaklinks=true]{hyperref}
\usepackage{amsmath,amssymb}
\usepackage{epsfig} 
\usepackage{graphicx}        
\usepackage{url}
\usepackage{color}
\usepackage{multirow}
\usepackage{placeins}
\usepackage[dvipsnames]{xcolor}
\usepackage{braket}
\usepackage{float}
\usepackage{slashed}
\hypersetup{
  colorlinks=true,
  linkcolor=red,
  filecolor=magenta,   
  urlcolor=blue,
  citecolor=blue,
}

\allowdisplaybreaks

\bibpunct{[}{]}{,}{n}{}{,}

\def\beq{\begin{equation}}
\def\eeq{\end{equation}}
\def\bea{\begin{eqnarray}}
\def\eea{\end{eqnarray}}
\makeatletter
\@addtoreset{equation}{section}

\begin{document}

\title{Explaining PTA Data with Inflationary GWs in a PBH-Dominated Universe}
\author{Satyabrata Datta}
\email{satyabrata.datta@saha.ac.in}
\affiliation{Saha Institute of Nuclear Physics, 1/AF, Bidhannagar, Kolkata 700064, India}
\begin{abstract}

We show that an ultralight primordial black hole (PBH) dominated phase makes blue-tilted inflationary gravitational waves (BGW) compatible with the recent detection of an nHz stochastic GW background by pulsar-timing arrays (PTAs),  for high reheating temperatures. This PBH-dominated phase suppresses the BGW spectrum via entropy dilution and generates a new GW spectrum from PBH density fluctuations.
This combined spectrum is detectable at ongoing and planned near-future GW detectors and exhibits a unique shape with a low-frequency peak explaining PTA data, a mid-range dip, and a sharp peak followed by a third peak at high-frequency. This distinctive shape sets it apart from spectra generated by other matter dominations or exotic physics. Therefore, while important for studying GWs in the nHz range, the recent PTA result also sets the stage for testing and constraining various well-studied mechanisms following a PBH domination, using low-frequency measurements and correlated observations of unique high-frequency GW spectral features.

\end{abstract}

\maketitle
\section{ Introduction} Recently, collaborations of pulsar-timing arrays (PTAs) such as NANOGrav, EPTA, and PPTA, along with InPTA and CPTA, have published their latest data. This data presents substantial evidence for a stochastic gravitational wave background (SGWB) at nHz frequencies \cite{NANOGrav:2023gor,EPTA:2023fyk,Reardon:2023gzh,Xu:2023wog}. A similar discovery, although with less statistical significance, has been present for the past two years, generating considerable interest within the scientific community \cite{EPTA:2021crs,NANOGrav:2020bcs,Goncharov:2021oub}. Intriguingly, this time, the signal displays the distinct angular correlations of pulsars, referred to as the quadrupolar Hellings-Downs curve \cite{Hellings:1983fr}, a feature that is specific to an SGWB. While the origins of such GWs are still uncertain, the favored power-law $\Omega_{\rm GW}\propto f^{1.8 \pm 0.6}$,  for instance, in the new NANOGrav data, does not rule out the straightforward GW-driven models of supermassive black hole binaries (SMBHBs) at $3\sigma$. Yet another intriguing prospect is to explore the GWs that originate from cosmological sources. In a companion theory paper \cite{NANOGrav:2023hvm}, the NANOGrav collaboration (we shall specifically focus on NANOGrav 15 yrs data \cite{NANOGrav:2023gor}; the results of other PTAs align well) has compiled a comprehensive list discussing numerous cosmological sources that are consistent with the data \footnote{Contrary to the 12.5 years of NANOGrav data, which was well-matched by stable cosmic strings \cite{Blasi:2020mfx,Ellis:2020ena,Samanta:2020cdk,Datta:2020bht,Samanta:2021mdm,Borah:2022iym}, the latest data seems to challenge this fit, suggesting that stable cosmic strings may not be the best explanation \cite{NANOGrav:2023gor}.}. Following this, various articles have explored these sources either within the framework of different cosmological models or by re-analyzing the fit to the new data, incorporating the results from other PTAs \cite{Ellis:2023tsl,Wang:2023len,Kitajima:2023cek,Franciolini:2023pbf,Megias:2023kiy,Fujikura:2023lkn,Han:2023olf,Zu:2023olm,Yang:2023aak,Guo:2023hyp,Shen:2023pan,Franciolini:2023wjm,Lambiase:2023pxd,Li:2023yaj,Han:2023olf,Zu:2023olm,Athron:2023mer,Babichev:2023pbf,Lazarides:2023ksx,Maji:2023fhv,Huang:2023chx,Jiang:2023gfe,Gangopadhyay:2023qjr,Nikandish:2023eak,Kawai:2023nqs,Frosina:2023nxu,Yi:2023tdk,Bhaumik:2023wmw,Wang:2023len,HosseiniMansoori:2023mqh,Yamada:2023thl,Ye:2023xyr,Fujikura:2023lkn,Madge:2023cak,Bringmann:2023opz,Depta:2023qst,Broadhurst:2023tus,Ge:2023rce,Oikonomou:2023qfz,Oikonomou:2023bli,Kitajima:2023vre,Eichhorn:2023gat,Broadhurst:2023tus,Unal:2023srk,DiBari:2023upq,Du:2023qvj,Antusch:2023zjk,Zhang:2023nrs,An:2023jxf,He:2023ado,Ellis:2023oxs,Kawasaki:2023rfx,Huang:2023zvs}. Inflationary GWs with a significant tensor blue-tilt, also known as Blue Tilted Gravitational Waves (BGWs), are a notable aspect that aligns well with both past and recent data \cite{EPTA:2023fyk,NANOGrav:2023gor,Vagnozzi:2020gtf,Bhattacharya:2020lhc,Kuroyanagi:2020sfw,Benetti:2021uea,Vagnozzi:2023lwo}; although, It’s important to mention that such BGWs can be generated in models that may not necessarily align with the conventional slow-roll inflation paradigm, see, e.g.,  \cite{Gruzinov:2004ty,Kobayashi:2010cm,Endlich:2012pz,Cannone:2014uqa,Ricciardone:2016lym,Cai:2014uka,Fujita:2018ehq,Mishima:2019vlh}. However, the range of parameters that allow for such a fit is, in fact, limited. This is primarily because GWs with large blue-tilt saturate big bang nucleosynthesis (BBN) bound on the effective number of neutrino species, disfavoring any post-inflationary cosmology founded on high reheating temperature ($T_{\rm RH}\gtrsim 10$ GeV) after inflation \cite{NANOGrav:2023hvm,Vagnozzi:2023lwo}. Nonetheless, if a non-standard matter epoch leads to entropy production between the reheating after inflation and the most recent radiation domination before the BBN \cite{Cyburt:2015mya}, BGWs get suppressed and provide a good fit to PTA data for high reheating temperatures. Now, contrary to the standard case, such a scenario allows the overall GW spectrum to span decades of frequencies with characteristic spectral features testable by high-frequency detectors, e.g., by the future LIGO run \cite{KAGRA:2021kbb,Peimbert:2016bdg, LIGOScientific:2016jlg}.

Building on the methodology outlined in the ref.\cite{Datta:2022tab,Datta:2023vbs}, we carried out a tomographic analysis of early matter domination (EMD) resulting from PBHs and their imprint on BGWs. While previous research has explored the impact of EMD induced by PBHs on the GW spectrum generated by cosmic strings \cite{Datta:2020bht,Samanta:2021mdm,Borah:2022iym,Borah:2022vsu,Borah:2023iqo}, no studies have specifically focused on BGWs and GWs originating from PBHs. Such ultralight PBH domination has been extensively studied within the context of Beyond Standard Model (BSM) physics scenarios, such as superheavy dark matter (DM) \cite{Samanta:2021mdm} and leptogenesis \cite{Datta:2020bht,Borah:2022iym}.

Motivated by these, we have undertaken a systematic investigation into the impact of a PBH-dominated epoch on BGWs, with a particular focus on explaining the results of the NANOGrav at low frequencies. We have demonstrated that the characteristic features of PBHs, specifically their mass and initial energy fraction, leave a mark on BGWs. This necessitates a larger initial fraction, which in turn leads to a higher entropy injection. This higher entropy injection suppresses the BGWs at high frequencies to ensure compliance with the constraints imposed by the LIGO and BBN.
Among the various mechanisms through which PBHs can generate GWs, we focus on the one induced by PBH density fluctuations \cite{Carr:2020gox,Inomata:2019ivs,Papanikolaou:2020qtd,Domenech:2020ssp,Domenech:2021wkk,Domenech:2021ztg}. While a requirement for a higher initial energy fraction of PBHs appears to conflict directly with GWs from density fluctuations saturating the constraints imposed by the LIGO and BBN, we have successfully identified a complementary region. This region not only addresses PTA data at low frequency and a peak at high frequency, but it also presents a unique, sharply peaked GW spectrum at mid-frequencies due to PBH density fluctuations. Thus, the overall spectrum introduces a rich phenomenology with characteristic features that can be explored in a complementary manner by future gravitational wave detectors.

The structure of the paper is as follows. In Section \ref{s2}, we provide a brief discussion on GWs from inflation with tensor blue tilt. Section \ref{s3} is dedicated to the dynamics of PBHs and the necessary components for studying their effect on BGWs. In Section \ref{s4}, we conduct a numerical analysis, prioritizing the satisfaction of NANOGrav data and examining PBH archaeology with BGWs. Section \ref{s5} discusses a unique possibility of detecting a complementary signal from BGWs and GWs resulting from PBH density fluctuations. Finally, in Section \ref{s6}, we conclude our findings.

\section{ Blue-tilted GWs from inflation}\label{s2}
One of the most plausible explanations for the origin of primordial GWs is cosmic inflation \cite{Guth:1980zm,Linde:1981mu}. In this section, we will briefly discuss how GWs are produced during inflation and how they travel through different cosmic eras until they reach the present day. GWs are described as a perturbation in the FLRW line element: 
\bea
ds^2=a(\tau)\left[-d\tau^2+(\delta_{ij}+h_{ij})dx^idx^j)\right],
\eea
with $\tau$ being the conformal time, and $a(\tau)$ the scale factor. The GWs are interpreted by the transverse and traceless ($\partial_ih^{ij}=0$, $\delta^{ij}h_{ij}=0$) part of the of the $3\times 3$ symmetric matrix $h_{ij}$. Since the GWs are too feeble, $|h_{ij}|\ll1$, the linearized evolution equation 
\bea
\partial_\mu(\sqrt{-g}\partial^\mu h_{ij})=16\pi a^2(\tau) \mathcal{\pi}_{ij}\label{lineq}
\eea
is sufficient for studying their propagation. The tensor part of the anisotropy stress, $\pi_{ij}$, coupled with $h_{ij}$ acts as an external source. It is useful to express $h_{ij}$ in the Fourier space: 
\bea
h_{ij}(\tau, \vec{x})=\sum_\lambda\int \frac{d^3\vec{k}}{(2\pi)^{3/2}} e^{i\vec{k}.\vec{x}}\epsilon_{ij}^\lambda(\vec{k})h_{\vec{k}}^\lambda(\tau),\label{fug}
\eea
where the index $\lambda=``+/-"$ denotes that the GWs have two polarisation states. The polarization tensors, in addition to being transverse and traceless, also fulfill the conditions: 
\begin{equation}
\begin{aligned}
{} &\text{(i)}\: \epsilon^{(\lambda)ij}(\vec{k})\epsilon_{ij}^{(\lambda^\prime)}(\vec{k})=2\delta_{\lambda\lambda^\prime}\\ 
&\text{(ii)} \:\epsilon^{(\lambda)}_{ij}(-\vec{k})=\epsilon^{(\lambda)}_{ij}(\vec{k}).
\end{aligned}
\end{equation}
Assuming that each polarization state evolves identically and isotropically, we can rename the notation by letting $h_{\vec{k}}^\lambda(\tau)$ as $h_{k}(\tau)$, where $k=|\vec{k}|=2\pi f$ with $f$ being the frequency of the GWs today at $a_0=1$. Taking into account the sub-dominant contribution from $\pi_{ij}$, the equation for the propagation of GWs in Fourier space can be written as
\bea
\ddot{h}_k+2\frac{\dot{a}}{a}\dot{h}_k+k^2h_k=0, \label{prpeq}
\eea
where the dot represents a derivative with respect to the conformal time. By utilizing Eq.\eqref{fug} and Eq.\eqref{prpeq}, we can compute the energy density of the GWs as \cite{WMAP:2006rnx}
\bea
\rho_{\rm GW}=\frac{1}{32\pi G}\int\frac{dk}{k}\left(\frac{k}{a}\right)^2T_T^2(\tau, k)P_T(k),\label{gw1}
\eea
where $T_T^2(\tau, k)=|h_k(\tau)|^2/|h_k(\tau_i)|^2$ is a transfer function that is derived from Eq.\eqref{prpeq}, with $\tau_i$ being an initial conformal time. The primordial power spectrum, $P_T(k)=\frac{k^3}{\pi^2}|h_k(\tau_i)|^2$ is linked to the inflation models with specific forms and is parametrised as a power-law given by
\bea
P_T(k)=r A_s(k_*)\left(\frac{k}{k_*}\right)^{n_T},
\eea

where $r\lesssim 0.06$ \cite{BICEP2:2018kqh} is the tensor-to-scalar-ratio,  $A_s \simeq 2\times 10^{-9}$ represents the scalar perturbation amplitude at the pivot scale $k_*=0.01\rm  Mpc^{-1}$ and the tensor spectral index is denoted by $n_T$. Interestingly, the standard slow-roll inflation models satisfy a consistency relation $n_T=-r/8$ \cite{Liddle:1993fq}, which results in slightly red-tilted GWs with $n_T\lesssim 0$. However, in this work, we have considered GWs with a significant blue tilt, where $n_T>0$, and we have assumed it to be constant throughout. The GW energy density, which is crucial for detection purposes, can be expressed as 
\bea
\Omega_{\rm GW}(k)=\frac{k}{\rho_c}\frac{d\rho_{\rm GW}}{dk},
\eea
where the quantity $\rho_c=3H_0^2/8\pi G$ with   $H_0\simeq 2.2 \times 10^{-4}~\rm Mpc^{-1}$ being the present-day Hubble constant. From Eq.\eqref{gw1}, the quantity $\Omega_{\rm GW}(k)$ is computed as 
\bea
\Omega_{\rm GW}(k)=\frac{1}{12H_0^2}\left(\frac{k}{a_0}\right)^2T_T^2(\tau_0,k)P_T(k),\text{ with} ~~\tau_0=1.4\times 10^4 {\rm ~Mpc}.\label{GWeq}
\eea

There have been various attempts to compute the transfer function analytically \cite{Seto:2003kc,Boyle:2005se,Nakayama:2008wy,Kuroyanagi:2008ye}, and one of the commonly used one for standard reheating is given by\cite{Nakayama:2009ce,Kuroyanagi:2014nba}
\bea
T_T^2(\tau_0,k)=F(k)T_1^2(\zeta_{\rm eq})T_2^2(\zeta_{R}),
\eea

where $F(k)$ reads
\bea
F(k)=\Omega_m^2\left( \frac{g_*(T_{k,\rm in})}{g_{*0}}\right)\left(\frac{g_{*s0}}{g_{*s}(T_{k,\rm in})}\right)^{4/3}\left(\frac{3j_1(k\tau_0)}{k\tau_0}\right)^2.\label{fuk}
\eea

Here $j_1(k\tau_0)$ is the spherical Bessel function, $\Omega_m=0.31$, $g_{*0}=3.36$, and $g_{*0s}=3.91$. The scale-dependent $g_{*0(s)}(T_{k,\rm in})$, used in Eq.\eqref{fuk} can be approximated analytically as \cite{Kuroyanagi:2014nba,Watanabe:2006qe,Saikawa:2018rcs}
\bea
g_{*0(s)}(T_{k,\rm in})=g_{*0}\left(\frac{A+{\rm tanh~k_1}}{A+1}\right)\left(\frac{B+{\rm tanh~k_2}}{B+1}\right),
\eea
where 
\bea
A=\frac{-1-10.75/g_{*0(s)}}{-1+10.75/g_{*0(s)}},~~B=\frac{-1-g_{max}/10.75 }{-1+g_{max}/10.75}, 
\eea
and 
\bea
k_1=-2.5~{\rm log}_{10}\left(\frac{k/2\pi}{2.5\times 10^{-12}{\rm Hz}}\right), \\
k_2=-2.0~{\rm log}_{10}\left(\frac{k/2\pi}{6.0\times 10^{-9}{\rm Hz}}\right).
\eea

The transfer functions are given by 
\bea
T_1^2(\zeta)=1+1.57\zeta+ 3.42 \zeta^2,\\
T_2^2(\zeta)=\left(1-0.22\zeta^{1.5}+0.65\zeta^2 \right)^{-1},
\eea
where $\zeta_i\equiv k/k_i$, with  $k_i$s being the wave number of the modes entering the horizon at different epochs and are derived as
\bea
k_{\rm eq}&=&7.1\times 10^{-2}\Omega_m h^2 {\rm Mpc^{-1}}.
\eea
In this paper, we shall use $k_* = 0.01$ Mpc$^{-1}$ and $h = 0.7$. 
It’s essential to highlight that a significant limitation to consider regarding the potential for BGWs is the $\Delta N_{\rm eff}$ bound from BBN, as well as the absence of any SGWB detection by LIGO \cite{Peimbert:2016bdg,LIGOScientific:2016jlg}. In the following section, we will demonstrate that if any late-time entropy production occurs through the evaporation of ultralight PBHs after significant PBH domination before reheating, it can considerably alter the transfer function during the PBH-dominated era and suppress the GW spectrum for modes that entered the horizon during the PBH-dominated phase. This could potentially ameliorate the constraints from BBN and LIGO.

\section{ Diluting BGWs through Entropy injection in PBH Domination}\label{s3}
 The dynamical evolution of energy densities of the black holes ($\rho_{\rm BH}$), and radiation ($\rho_{\rm R}$) is governed by the following Friedmann equations
 \cite{Datta:2020bht,Masina:2020xhk}:
\bea
\frac{d\rho_{R}}{dz}+\frac{4}{z}\rho_R=0,\label{den1}\\
\frac{d\rho_{\rm BH}}{dz}+\frac{3}{z}\frac{H}{\tilde{H}}\rho_{\rm BH}-\frac{\dot{M}_{\rm BH}}{M_{\rm BH}}\frac{1}{z\tilde{H}}\rho_{\rm BH}=0,\label{den2}
\eea
where $H$ is the Hubble parameter and $z=T_{\rm Bf}/T$. The scale factor $a$ and the quantity $\tilde{H}$ evolve as
\bea
\tilde{H}=\left(H+\mathcal{K}\right),~
\frac{da}{dz}=\left(1-\frac{\mathcal{K}}{\tilde{H}}\right)\frac{a}{z},
\label{temvar}
\eea
where $\mathcal{K}=\frac{\dot{M}_{\rm BH}}{M_{\rm BH}}\frac{\rho_{\rm BH}}{4\rho_{\rm R}}$. In the derivation of Eq.\eqref{den1}-Eq.\eqref{temvar}, we have made the assumption that entropy ($g_{*s}$) and the energy ($g_{*\rho}$) degrees of freedom are equal and constant. If the initial energy fraction of PBHs, denoted as $\beta\equiv \frac{\rho_{\rm BH}(T_{\rm Bf})}{\rho_{\rm R}(T_{\rm Bf})}$, surpasses a critical value of $\beta_c\equiv \gamma^{-1/2}\sqrt{\left({\mathcal{G}g_{*B} (T_{\rm BH})}/{10240\pi} \right)}\frac{M_{\rm Pl}}{M_{\rm BH}}$, it can result in early matter domination. For a given $\beta$, the above equations can be solved numerically to determine the temperatures of PBH domination and evaporation, as well as the entropy production $\Delta_{\rm PBH}=\tilde{S}_{2}/\tilde{S}_{1}$, where $\tilde{S}_{1,\:(2)}\propto a_{1,\:(2)}^3/z_{1,\:(2)}^3$ is the total entropy before (after) the PBH evaporation. We will ultimately utilize analytical expressions that are in close agreement with numerical results and can be approximated accordingly \cite{Borah:2022iym}
 \bea
\Delta_{\rm PBH} \simeq 233 \beta \left( \frac{M_{\rm BH}}{M_{\rm Pl}}\right)\left( \frac{\gamma}{g_{*B}(T_{\rm BH})\mathcal{G}}\right)^{1/2},\label{entr}
 \eea
 where $\gamma\simeq 0.2$ is the formation efficiency of PBHs, $g_{*B}\simeq 100$ is the no of relativistic d.o.f. below the Hawking temperature $T_{\rm BH}$, and $\mathcal{G}\simeq 3.8$ is the greybody factor.

 In the present scenario where late-time entropy production takes place after reheating through PBH evaporation, the background evolution follows the sequence of MD (inflation-dominated) $\rightarrow$ RD $\rightarrow$ MD (PBH-dominated) $\rightarrow$ RD. In such a
case, the transfer function is defined by 
\bea
T_T^2(\tau_0,k)=F(k)T_1^2(\zeta_{\rm eq})T_2^2(\zeta_{\text{ev}})T_3^2(\zeta_{\text{dom}})T_2^2(\zeta_{R}),\label{transfer}
\eea

where $T_3^2(\zeta)$ is the transfer function corresponding to the transition from the first RD phase to the PBH-dominated phase and is given by
\bea
T_3^2(\zeta)=1+0.59\zeta+0.65 \zeta^2,
\eea
and the modes $k_i$'s given by
\bea
k_{\text{ev}}&=&1.7\times 10^{14}\left(\frac{g_{*s}(T_\text{ev})}{106.75}\right)^{1/6}\left(\frac{T_\text{ev}}{10^7 \rm GeV}\right){\rm Mpc^{-1}},\label{kev} \\
k_{\text{dom}}&=&1.7\times 10^{14} \left(\frac{g_{*s}(T_\text{dom})}{106.75}\right)^{1/6}\left(\frac{T_\text{dom}}{10^7 \rm GeV}\right){\rm Mpc^{-1}},\label{kdom}
\eea
and 
\bea
k_{\rm R}=1.7\times 10^{14}\Delta_{\rm PBH}^{-1/3}\left(\frac{g_{*s}(T_{\rm R})}{106.75}\right)^{1/6}\left(\frac{T_{\rm R}}{10^7 \rm GeV}\right){\rm Mpc^{-1}} \label{krh}
\eea
which reenters the horizon at PBH evaporation temperature ($T_{\rm ev}$), PBH domination temperature ($T_{\rm dom}$), and inflationary reheating temperature ($T_{\rm R}$), respectively where \cite{Datta:2020bht}
\bea
T_{\rm ev}=\left(\frac{45 M_{\rm Pl}^2}{16\pi^3 g_*(T_{\rm ev})\tau^2}\right)^{1/4},\label{ev}
\eea
and 
\bea
T_{\rm dom}=\beta T_{\rm Bf},
\eea
where $\tau$, and $T_{\rm Bf}$ correspond to the BH lifetime and BH formation temperature respectively, and are given by
\bea
\tau= \frac{10240 \pi M_{\rm BH}^3}{\mathcal{G} g_{*B}(T_{\rm BH})M_{\rm Pl}^4},\label{life}
\eea
and 
\bea
T_{\rm Bf}=\left(  \frac{45\gamma^2}{16\pi^3 g_*(T_{\rm Bf})}\right)^{1/4} \left(\frac{M_{\rm Pl}}{M_{\rm BH}} \right)^{1/2} M_{\rm Pl} .
\eea
Note also that, by construction, the formation of BHs occurs after the initial inflationary reheating, i.e. $T_{\rm Bf}\lesssim T_{\rm R}$. For simplicity, we shall use $T_{\rm R}=T_{\rm Bf}$ throughout the rest of our discussion.

\section{A fit to the NANOGrav data in a PBH-dominated scenario and PBH archaeology with BGWs} \label{s4}
Taking into account the aforementioned equations and utilizing $\Delta_{\rm PBH}$ from Eq.\eqref{entr}, we evaluate Eq.\eqref{GWeq} to get the GW spectrum and to fit the NANOGrav data, while considering two additional constraints. {\bf I)} The LIGO O3 constraint on SGWB, which can be approximately expressed as $\Omega_{\rm GW} (25 \text{\:Hz})h^2\leq 2.2\times 10^{-9}$ \cite{KAGRA:2021kbb}, and {\bf II)} BBN bound on the effective number of neutrino species: $\int_{f_{\rm low}}^{f_{\rm high}}f^{-1}df \Omega_{\rm GW}(f)h^2\lesssim 5.6\times 10^{-6}\Delta N_{\rm eff}$, where $\Delta N_{\rm eff}\lesssim 0.2$ \cite{Planck:2018vyg}. The lower limit of the integration is associated with the frequency that entered the horizon during the BBN epoch. Conversely, the Hubble rate at the end of inflation sets the upper limit: $f_{\rm high}=a_{\rm end} H_{\rm end}/2\pi$. For numerical calculations, $f_{\rm high}\simeq 10^5$ Hz is sufficient as the spectrum falls and the integration saturates.
\begin{figure}
\centering
 \includegraphics[scale=.6]{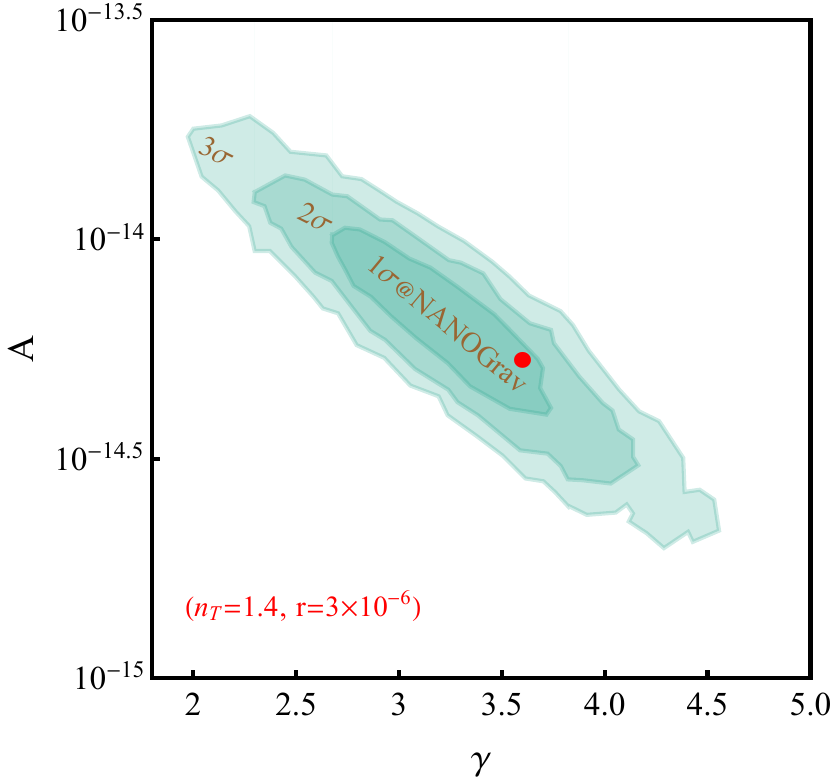}   
 	\caption{Top: The red $\bullet$ represents the amplitude $A$ and the spectral index $\gamma$ corresponding to a benchmark ($n_T$, $r$)$\equiv$ ($1.4$, $3\times10^{-6}$).  NANOGrav $1,2,3\sigma$ regions are shown by the cyan shades.  }\label{fig1}
 \end{figure}

We follow the NANOGrav parametrization for the GW energy density to conduct a power-law fit to the new data within the frequency range $f\in\left[10^{-9}~{\rm Hz},f_{\rm yr}\right]$. The  parametrization reads
\begin{equation}
\Omega_{\rm GW}(f)=\Omega_{\rm yr}\left(\frac{f}{f_{\rm yr}}\right)^{(5-\gamma)}\label{pl}
\end{equation}
with $\Omega_{\rm yr}=\frac{2\pi^2}{3H_0^2}A^2 f_{\rm yr}^2$ and $f_{\rm yr}=1{\rm yr}^{-1}\simeq 32$ nHz. To fit the data, one must compare Eq.(\ref{GWeq}) and Eq.(\ref{pl}), then extract the values of the amplitude $A$ and the spectral index  $\gamma$ that fall within the $1,\:2,\:3\sigma$ contours as reported by the NANOGrav \cite{NANOGrav:2023gor}. In Fig.\ref{fig1}, we present a particular benchmark where $n_T=1.4$ and $r=3\times10^{-6}$ for which the BGW spectrum intersects with the 1$\sigma$ range of NANOGrav. For this specific benchmark point, the parameters that are vital to PBH dynamics, namely $\beta$ and $M_{\rm BH}$, have been adjusted freely. Due to this flexibility, for a wide range of $n_T$ and $r$, the BGW spectrum, which was previously ruled out by stringent constraints from LIGO and BBN at higher frequencies, can now be permitted by choosing appropriate values for $\beta$ and $M_{\rm BH}$. To provide a quantitative understanding of the impact of PBH domination, we consider the same benchmark point $n_T=1.4$ and $r=3\times10^{-6}$ as a reference.

\begin{figure}
\centering
\includegraphics[scale=.57]{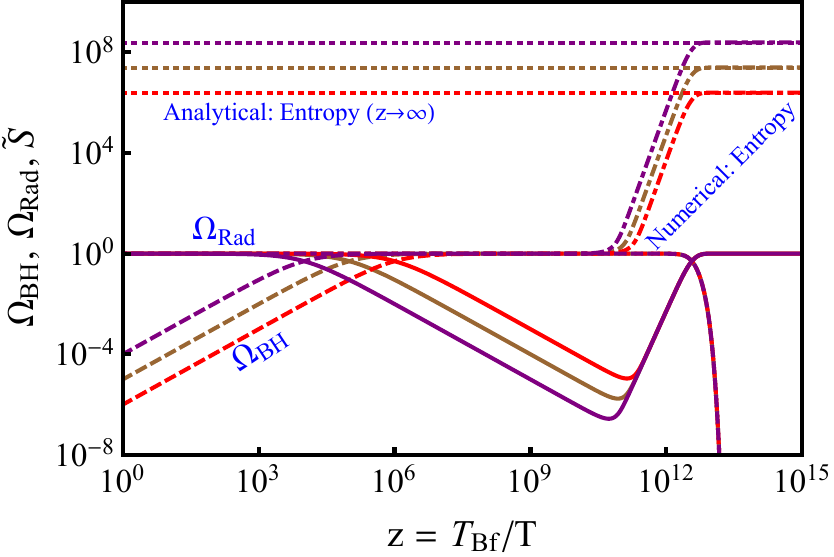}  \includegraphics[scale=.57]{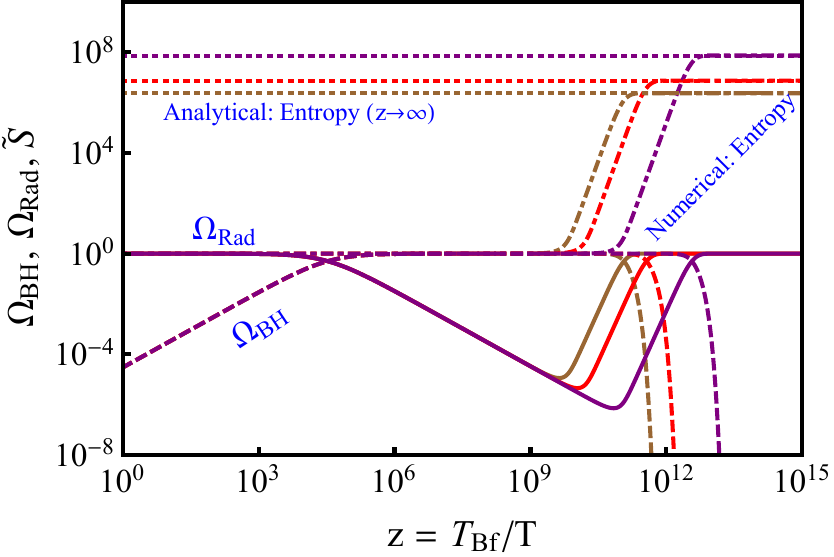} \\
 \includegraphics[scale=.57]{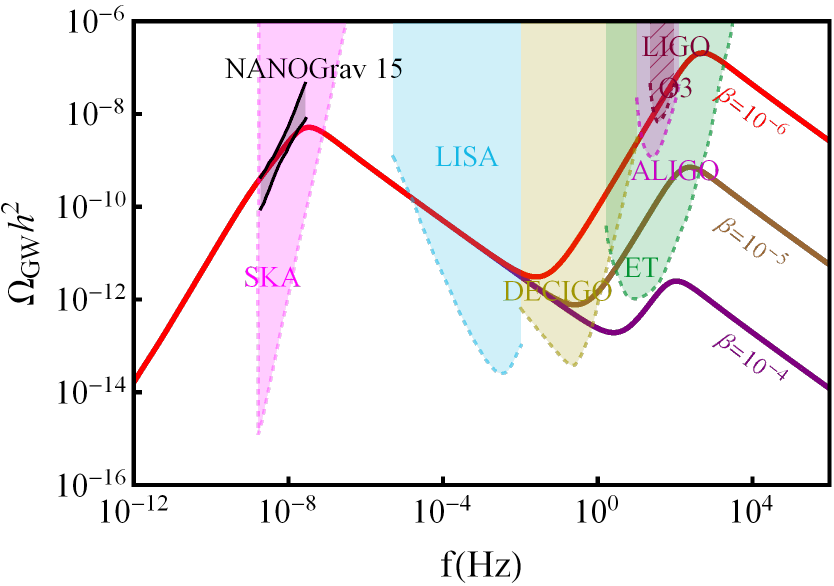} \includegraphics[scale=.57]{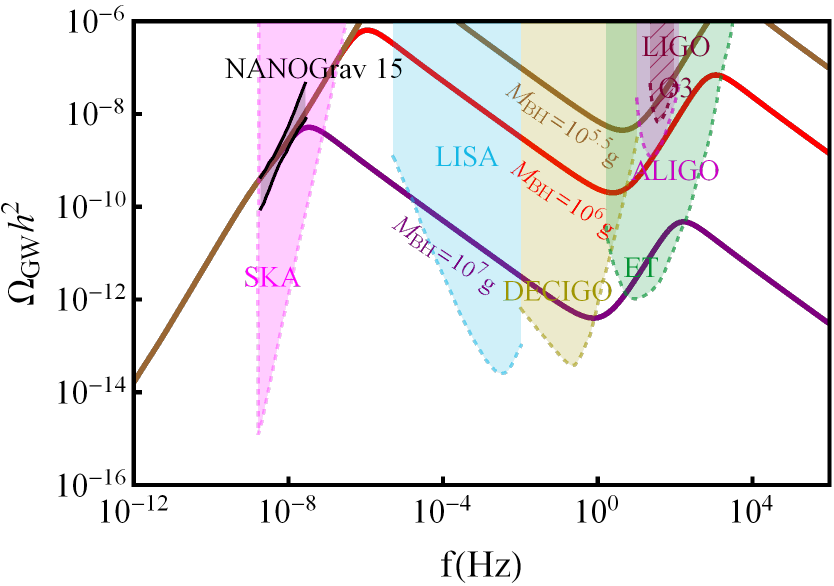}   
 	\caption{Top-left: Evolution of the radiation energy density (solid), PBH energy density (dashed), and the total entropy (dot-dashed) with the inverse temperature. The benchmark points are represented in different colors: red for $\beta=10^{-6}$, brown for $\beta=10^{-5}$, and purple for $\beta=10^{-4}$. In all these cases, the PBH mass, $M_{\rm BH}$ is fixed at $10^7$ grams. Top-right: Evolution of the same quantities, but for different PBH masses and fixed initial energy fraction. Bottom-left/right: The BGW spectra comply with the quantities that determine the nature of the corresponding plots on the top.}\label{fig2}
 \end{figure}

In the top/bottom left plot of Fig.\ref{fig2}, we varied $\beta$ for three distinct values: $10^{-6}$ (depicted in red), $10^{-5}$ (depicted in brown), and $10^{-4}$ (depicted in purple), while keeping $M_{\rm BH}=10^7$ grams fixed. Conversely, in the top/bottom right plot of Fig.\ref{fig2}, we varied $M_{\rm BH}$ for three different values: $10^{5.5}$ grams (shown in brown), $10^6$ grams (shown in red), and $10^{7}$ grams (shown in purple) with $\beta=3\times10^{-5}$ held constant.   

Interestingly, it is possible to find the analytical expressions of three turning point frequencies as a function of $\beta$ and $M_{\rm BH}$ (c.f. Eq.\eqref{kev}, \eqref{kdom}, \eqref{krh}) and these are shown as follows: 
\bea
f_{\rm ev}=4.88\times 10^{-10} M_{\rm Pl}\left( \frac{g_{\rm *s}(T_{\rm ev})}{106.75}\right)^{-1/12}\left( \frac{g_{\rm *B}}{100}\right)^{1/2}\left( \frac{M_{\rm Pl}}{M_{\rm BH}}\right)^{3/2},\label{flow}\\
f_{\rm dom}=2.04\times 10^{-17} M_{\rm Pl}\left( \frac{\beta}{10^{-8}}\right)\left( \frac{g_{\rm *B}}{100}\right)^{-1/4}\left( \frac{g_{\rm *s}(T_{\rm dom})}{106.75}\right)^{1/6}\left(\frac{M_{\rm Pl}}{M_{\rm BH}} \right)^{1/2},\label{fdip}\\
f_{\rm R}=5.42\times 10^{-7} M_{\rm Pl}\left( \frac{\beta}{10^{-8}}\right)^{-1/3}\left( \frac{g_{\rm *B}}{100}\right)^{-1/12}\left( \frac{g_{\rm *s}(T_{\rm R})}{106.75}\right)^{1/6}\left(\frac{M_{\rm Pl}}{M_{\rm BH} }\right)^{5/6}.\label{fhigh}
\eea

From the inspection of the subplots in Fig.\ref{fig2} and Eqs.\eqref{GWeq}, \eqref{transfer}, \eqref{flow}, \eqref{fdip}, \eqref{fhigh}, we can extract the following pieces of information, {\bf I)} The variation of $\beta$ has a significant impact on the second peak and the mid-frequency dip, i.e., Eq.\eqref{fhigh} and \eqref{fdip}, respectively. For larger values of $\beta$, the high-frequency peak shifts towards lower frequency according to Eq.\eqref{fhigh}. Additionally, the dip shifts towards higher frequency as per Eq.\eqref{fdip}. Furthermore, with increasing $\beta$, the BGW amplitude at high frequency gets suppressed while the amplitude at low frequency remains unaffected. The rationale for this can be readily comprehended by examining the top-left plot of Fig.\ref{fig2}. As $\beta$ increases, the strength of matter domination will be much stronger, leading to a larger entropy injection that dilutes BGWs at high frequency. The evaporation temperature (see Eq.\eqref{ev}), which is not influenced by $\beta$, indicates that the BGW amplitude at low-frequency peak (c.f. Eq.\eqref{transfer}), which depends on the evaporation temperature, remains unchanged. {\bf II)} Altering $M_{\rm BH}$ impacts the entire BGW spectrum and all the turning point frequencies. As can be observed from the top-right plot of Fig.\ref{fig2}, an increase in $M_{\rm BH}$ extends the lifetime of PBHs (See Eq.\eqref{life}). Consequently, this leads to a larger entropy injection, resulting in a more diluted BGW spectrum. \\

We end this section by highlighting a potential challenge: in the future, if GW detectors identify such inflationary double-peak GWs, we may not be able to distinguish between ultralight PBH domination and other intermediate matter domination scenarios. However, there is an intriguing solution to this issue. For certain PBH model parameters, as we will demonstrate next, they can generate their own GWs from PBH density fluctuations.
\section{Detection Prospects for the Unique GW Spectrum(BGW+PBH density fluctuations) and Realistic BSM Physics Scenarios}\label{s5}
It’s noteworthy to state that ultralight PBHs play multiple roles in the generation of GWs. For instance, the
initial curvature perturbations that lead to the formation of PBHs also give rise to GWs (see, e.g., \cite{Saito:2008jc,Cai:2018dig,Inomata:2020lmk}), PBHs emit gravitons, which constitute high-frequency GWs \cite{Anantua:2008am}, PBHs also merge, releasing GWs
\cite{Hooper:2020evu}, Lastly, the inhomogeneous distribution of PBHs, which results in density fluctuations, triggers the production of GWs \cite{Papanikolaou:2020qtd,Domenech:2020ssp,Domenech:2021wkk}. In this section, we will concentrate on the last point mentioned above.
As recently highlighted in ref.\cite{Papanikolaou:2020qtd} and further elaborated in refs. \cite{Domenech:2020ssp,Domenech:2021wkk}, it’s observed that PBHs, immediately after their formation, are distributed randomly in space following Poisson distribution. Hence, even though the PBH gas, on average, behaves like pressure-less dust, the uneven spatial distribution results in density fluctuations, which are inherently isocurvature. When PBHs start to dominate the energy density of the Universe, the isocurvature component transitions into curvature perturbations, which subsequently give rise to secondary GWs. Due to the significant density fluctuations at smaller scales (equivalent to the average separation of PBHs at $T_{\rm Bf}$), substantial GWs are generated.

These GWs are further amplified due to the nearly instantaneous evaporation of PBHs. The amplitude of such induced GWs in the present day is given by\footnote{Note that the amplitude of the induced GWs is inherently independent of the history of PBH formation and, by design, does not rely on any non-standard inflation dynamics \cite{Domenech:2020ssp,Domenech:2021wkk}.}
\bea
\Omega_{\rm GW}^{\rm PBH}(t_0,f)\simeq \Omega^{p}_{\rm GW} \left( \frac{f}{f_p}\right)^{11/3} \Theta (f_{p}-f),
\eea
where the peak amplitude is given by
\bea
\Omega^{p}_{\rm GW}\simeq 2\times 10^{-6}\left(\frac{\beta}{10^{-8}}\right)^{16/3} \left(\frac{M_{\rm BH}}{10^7 g}\right)^{34/9},\label{peakamp}
\eea
and the peak frequency is given by
\bea
f_p\simeq 1.7\times 10^3 \text{\:Hz}\left( \frac{M_{\rm BH}}{10^4 g}\right)^{-5/6}.
\eea
The ultraviolet cutoff represented by the $\Theta$ function, which is on par with the frequency that corresponds to the comoving scale, signifies the average separation of PBHs at their time of formation.

\begin{figure}
\centering
 \includegraphics[scale=.55]{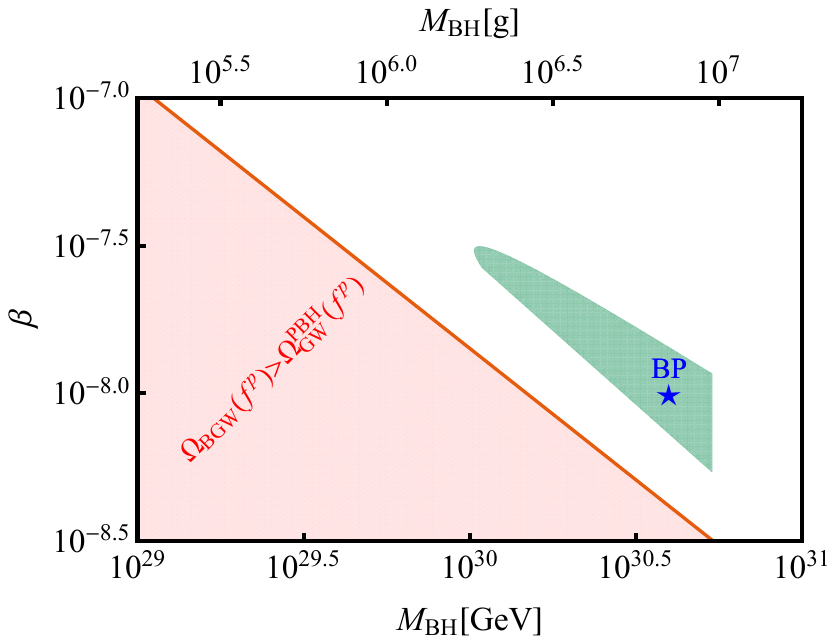} 
 \includegraphics[scale=.55]{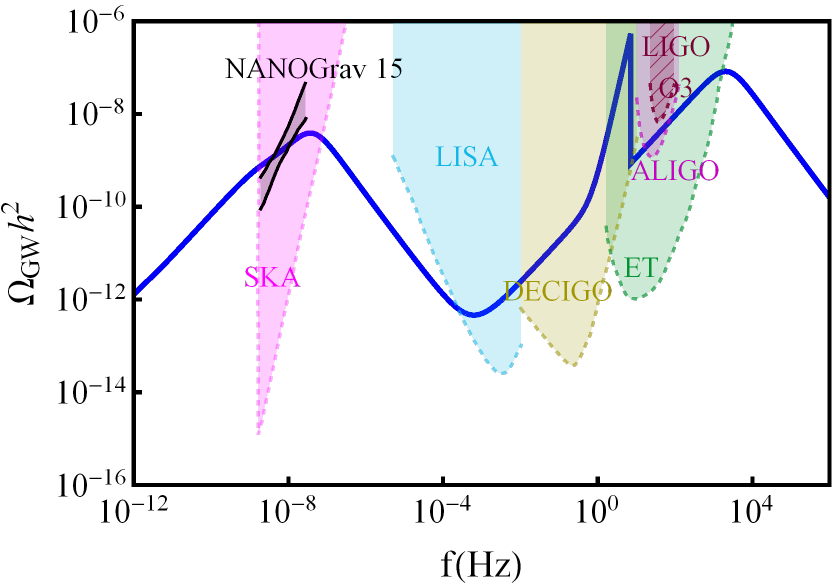}  
 	\caption{Left: Allowed region of combined GW spectra from BGW and PBH density fluctuations for $n_T=0.9$ and $r=0.06$, Right: Combined GW spectrum subjected to the benchmark point ($M_{\rm BH}=7.08\times 10^{6}$gram, $\beta=10^{-8}$)}\label{fig3}
 \end{figure}

As previously demonstrated, PBHs with $\beta > \beta_c$ have the potential to cause spectral distortion
in the BGWs from inflation due to the domination and subsequent evaporation of PBHs. Concurrently, this also gives rise to a new source of GWs originating from density fluctuations leading to
a complex GW spectral shape characterized by three distinct peak features. The left plot in Fig.\ref{fig3} shows a red-shaded region that is excluded due to the condition $\Omega_{\rm BGW}(f^p)> \Omega_{\rm GW}^{\rm PBH}(f^p)$. This implies that the characteristic peak from PBH density fluctuations will always be suppressed below the BGW spectrum, rendering the PBH scenario indistinguishable from any other intermediate matter domination. We present an allowed parameter space on the $\beta$-$M_{\rm BH}$ plane for a benchmark point $n_T=0.9$ and $r=0.06$. Although this does not provide the best fit to the NANOGrav data, it fits the data at $2\sigma$. The green region is allowed, satisfying both the required frequency and amplitude by NANOGrav and the bound on SGWB from LIGO and BBN. Interestingly, throughout the entire parameter space, the condition $\Omega_{\rm BGW}(f^p)< \Omega_{\rm GW}^{\rm PBH}(f^p)$ holds true. This allows for the possibility of a characteristic GW signal from PBHs that exhibits three peaks. For this specific benchmark point, we have also ensured that the inequality $f_{\rm dom}<f^p<f_{\rm R}$ holds true. An important point to note is that while the parameters $n_T$ and $r$ can be varied to identify new allowed regions that fit the NANOGrav data more exhaustively, there is no qualitative difference. The allowed values of $n_T$ and $r$ are close to the presented benchmark. This is because, as seen in the bottom left plot of Fig.\ref{fig2}, higher $\beta$ values are needed to suppress the BGW spectrum at higher frequencies to satisfy the BBN and LIGO constraints. However, from Eq.\eqref{peakamp}, it can be seen that the GW peak amplitude from PBH density fluctuations increases with $\beta$. These two features, in conjunction with the need for consistency with the NANOGrav data, strongly constrain the parameter space.

Considering the broad overview provided earlier, we can highlight how such a spectrum could serve as an indicator for many BSM models in particle physics. Firstly, while such light PBHs may ultimately evaporate, they could generate relics that are either stable or unstable. A stable relic could potentially represent dark matter, while an unstable particle, such as right-handed neutrinos in the seesaw mechanism, that emerges from PBH could potentially trigger baryogenesis via leptogenesis (see, e.g., refs.\cite{Fujita:2014hha,Lennon:2017tqq,Domenech:2023mqk,Franciolini:2023osw}). Moreover, based on the mass and initial energy fraction, a universe dominated by PBHs can modify the standard parameter space of numerous particle physics models, such as dark matter models \cite{Gondolo:2020uqv}. Therefore, PBHs (their mass and initial energy fraction) serve as a gateway linking GWs to the parameters of high-energy particle physics models. This further demands a unified investigation of these models using both GWs and conventional particle physics experiments which is beyond the scope of this work. To conclude, we would like to emphasize that one of the main motivations of this paper is to show that the effects of an EMD caused by the standard long-lived fields or some additional new physics \cite{Asaka:2020wcr,Datta:2022tab,Borah:2023sbc} and PBHs could be very different. Specifically, distorting BGWs with EMD leads to a low-frequency peak followed by a dip and a high-frequency peak in both scenarios. However, in the case of EMD from PBH, depending on the initial energy fraction of PBHs, a third distinct sharp peak might appear at mid-frequency. This sets the PBH scenarios apart. Furthermore, the impacts of ultra-light PBHs, which are highly intriguing to investigate, for instance, in the context of dark matter production, baryogenesis, and axions, could make these spectral features in the GWs a unique indicator of these models, regardless of the strength of the particle physics coupling.
\section{Conclusion}\label{s6}
We have discussed a unique framework that uses NANOGrav PTA data, interpreted as a stochastic gravitational wave background (SGWB) from inflation, to investigate an epoch dominated by primordial black holes (PBH). With an appropriate choice of PBH parameters, a PBH-dominated epoch can accommodate almost any value of the inflationary parameters, fitting the GW spectrum at the NANOGrav frequency while also satisfying the stringent high-frequency constraints from LIGO and BBN.
The presence of any significant intermediate matter domination (here PBH domination) after inflation results in a double-peaked GW spectrum with a dip in between, while the introduction of ultralight PBHs contributes an additional GW spectrum from density fluctuations. The combined GW spectrum exhibits a unique shape, characterized by a low-frequency peak that explains the PTA data, a dip in the middle, and a sharp tilted peak, followed by a third high-frequency peak. Such a combined spectrum, along with the PTA fit and constraints from LIGO and BBN, predicts a highly constrained PBH parameter space that remains distinguishable from any other intermediate matter domination scenario. In addition to the distinct features that can be verified in GW detectors, the framework discussed in our work can also lead to a plethora of beyond the Standard model phenomenological implications, such as the production of dark matter from PBH evaporation and high-scale leptogenesis, among others.   

\section*{acknowledgement}
The author would like to thank Rome Samanta for useful discussions and careful reading of the manuscript.

\bibliographystyle{JHEP}
\bibliography{main.bib}
\end{document}